# Fundamental solution of Fokker - Planck equation

*Igor A. Tanski*
*121242, Ivanteevka Moscow region (Russia)*
*Centralny projezd, 4/69*
*tanski@protek.ru*

ZAO CV Protek

*ABSTRACT*

Fundamental solution of Fokker - Planck equation is built by means of the Fourier transform method. The result is checked by direct calculation.

We see from recent publications (ref. [1-5]) that the Fokker - Planck equation is an object of steady investigations. It is a very useful modelling tool in many fields. Therefore it is desirable to have exact expression for fundamental solution.

The object of our considerations is a special case of Fokker - Planck equation, which describes evolution of 3D continuum of non-interacting particles imbedded in a dense medium without outer forces. The interaction between particles and medium causes combined diffusion in physical space and velocities space. The only force, which acts on particles, is damping force proportional to velocity.

For this special case the Fokker - Planck equation is:

$$\frac{\partial n}{\partial t} + v_x \frac{\partial n}{\partial x} + v_y \frac{\partial n}{\partial y} + v_z \frac{\partial n}{\partial z} - \alpha \left( \frac{\partial}{\partial v_x}(v_x n) + \frac{\partial}{\partial v_y}(v_y n) + \frac{\partial}{\partial v_z}(v_z n) \right) = k \left( \frac{\partial^2 n}{\partial v_x^2} + \frac{\partial^2 n}{\partial v_y^2} + \frac{\partial^2 n}{\partial v_z^2} \right) \quad (1)$$

where

$n = n(t, x, y, z, v_x, v_y, v_z)$ - density;

$t$ - time variable;

$x, y, z$ - space coordinates;

$v_x, v_y, v_z$ - velocities;

$\alpha$ - coefficient of damping;

$k$ - coefficient of diffusion.

We shall search solution of equation (1) for unlimited space

$$-\infty < x < +\infty, -\infty < y < +\infty, -\infty < z < +\infty, \quad (2)$$

$$-\infty < v_x < +\infty, -\infty < v_y < +\infty, -\infty < v_z < +\infty.$$

Let us denote by $n_0$ initial density

$$n_0(x, y, z, v_x, v_y, v_z) = n(0, x, y, z, v_x, v_y, v_z). \tag{3}$$

We shall use the Fourier transform method. Let us denote by $N$ Fourier transform of density

$$N = N(t, p_x, p_y, p_z, q_x, q_y, q_z) = \tag{4}$$

$$\frac{1}{(2\pi)^6} \int_{-\infty}^{\infty}\int_{-\infty}^{\infty}\int_{-\infty}^{\infty}\int_{-\infty}^{\infty}\int_{-\infty}^{\infty}\int_{-\infty}^{\infty} \exp(-i(xp_x + yp_y + zp_z + v_x q_x + v_y q_y + v_z q_z)) n\, dx\, dy\, dz\, dv_x\, dv_y\, dv_z.$$

and by $N_0$ - Fourier transform of initial density:

$$N_0(p_x, p_y, p_z, q_x, q_y, q_z) = N(0, p_x, p_y, p_z, q_x, q_y, q_z) \tag{5}$$

where $p_x, p_y, p_z$ - space coordinates momentum variables;

$q_x, q_y, q_z$ - velocities momentum variables.

Multiplying (1) by $\exp(-i(xp_x + yp_y + zp_z + v_x q_x + v_y q_y + v_z q_z))$ and integrating over whole space and velocities, we obtain

$$\frac{\partial N}{\partial t} + (\alpha q_x - p_x)\frac{\partial N}{\partial q_x} + (\alpha q_y - p_y)\frac{\partial N}{\partial q_y} + (\alpha q_z - p_z)\frac{\partial N}{\partial q_z} = -k\left(q_x^2 + q_y^2 + q_z^2\right)N. \tag{6}$$

The equation (6) is a linear differential equation of first order, so it can be solved by the method of characteristics. Relations on the characteristics are

$$dt = \frac{dp_x}{0} = \frac{dp_y}{0} = \frac{dp_z}{0} = \frac{dq_x}{\alpha q_x - p_x} = \frac{dq_y}{\alpha q_y - p_y} = \frac{dq_z}{\alpha q_z - p_z} = -\frac{dN/N}{k\,(q_x^2 + q_y^2 + q_z^2)}. \tag{7}$$

First three equations (7) have three integrals:

$$p_x = const; \tag{8}$$

$$p_y = const;$$

$$p_z = const.$$

If we combine (8) with next three equations (7), we get three further integrals

$$(\alpha q_x - p_x)\, e^{-\alpha t} = const; \tag{9}$$

$$(\alpha q_y - p_y)\, e^{-\alpha t} = const;$$

$$(\alpha q_z - p_z)\, e^{-\alpha t} = const.$$

We solve (9) for $q_x, q_y, q_z$ and obtain

$$q_x = \frac{p_x}{\alpha} + e^{\alpha t}\left(q_{x0} - \frac{p_x}{\alpha}\right); \tag{10}$$

$$q_y = \frac{p_y}{\alpha} + e^{\alpha t}\left(q_{y0} - \frac{p_y}{\alpha}\right);$$

$$q_z = \frac{p_z}{\alpha} + e^{\alpha t}\left(q_{z0} - \frac{p_z}{\alpha}\right).$$

On the other hand, if we solve (9) for $q_{x0}, q_{y0}, q_{z0}$, we obtain

$$q_{x0} = \frac{p_x}{\alpha} + e^{-\alpha t}\left(q_x - \frac{p_x}{\alpha}\right); \qquad (11)$$

$$q_{y0} = \frac{p_y}{\alpha} + e^{-\alpha t}\left(q_y - \frac{p_y}{\alpha}\right);$$

$$q_{z0} = \frac{p_z}{\alpha} + e^{-\alpha t}\left(q_z - \frac{p_z}{\alpha}\right).$$

To get the last integral of (7), we replace current velocities momentum variables $q_x, q_y, q_z$ by initial velocities momentum variables $q_{x0}, q_{y0}, q_{z0}$ in $q_x^2 + q_y^2 + q_z^2$

$$q_x^2 + q_y^2 + q_z^2 = \frac{1}{\alpha^2}\left(p_x^2 + p_y^2 + p_z^2\right) + \qquad (12)$$

$$+ \frac{2}{\alpha} e^{\alpha t}\left(p_x(q_{x0} - \frac{p_s}{\alpha}) + p_y(q_{y0} - \frac{p_s}{\alpha}) + p_z(q_{z0} - \frac{p_s}{\alpha})\right) +$$

$$+ e^{2\alpha t}\left((q_{x0} - \frac{p_x}{\alpha})^2 + (q_{y0} - \frac{p_y}{\alpha})^2 + (q_{z0} - \frac{p_z}{\alpha})^2\right).$$

Integrating (12) in $t$, we obtain the last integral

$$ln(N) + k\left[\frac{t}{\alpha^2}(p_x^2 + p_y^2 + p_z^2) + \right. \qquad (13)$$

$$+ \frac{2}{\alpha^2} e^{\alpha t}\left(p_x(q_{x0} - \frac{p_x}{\alpha}) + p_y(q_{y0} - \frac{p_y}{\alpha}) + p_z(q_{z0} - \frac{p_z}{\alpha})\right) +$$

$$\left. + \frac{e^{2\alpha t}}{2\alpha}\left((q_{x0} - \frac{p_x}{\alpha})^2 + (q_{y0} - \frac{p_y}{\alpha})^2 + (q_{z0} - \frac{p_z}{\alpha})^2\right)\right] = const.$$

We replace in (13) initial values of velocities momentum variables by their current values

$$ln(N) + k\left[\frac{t}{\alpha^2}(p_x^2 + p_y^2 + p_z^2) + \right. \qquad (14)$$

$$+ \frac{2}{\alpha^2}\left(p_x(q_x - \frac{p_x}{\alpha}) + p_y(q_y - \frac{p_y}{\alpha}) + p_z(q_z - \frac{p_z}{\alpha})\right) +$$

$$\left. + \frac{1}{2\alpha}\left((q_x - \frac{p_x}{\alpha})^2 + (q_y - \frac{p_y}{\alpha})^2 + (q_z - \frac{p_z}{\alpha})^2\right)\right] = const.$$

To determine the constant term in (14), we write the same expression for initial values and equate both expressions

$$ln(N) + k\left[\frac{t}{\alpha^2}(p_x^2 + p_y^2 + p_z^2) + \right. \qquad (15)$$

$$+ \frac{2}{\alpha^2}\left(p_x(q_x - \frac{p_x}{\alpha}) + p_y(q_y - \frac{p_y}{\alpha}) + p_z(q_z - \frac{p_z}{\alpha})\right) +$$

$$+ \frac{1}{2\alpha}\left((q_x - \frac{p_x}{\alpha})^2 + (q_y - \frac{p_y}{\alpha})^2 + (q_z - \frac{p_z}{\alpha})^2\right)\right] =$$

$$ln(N_0) + k\left[\frac{2}{\alpha^2}\left(p_x(q_{x0} - \frac{p_x}{\alpha}) + p_y(q_{y0} - \frac{p_y}{\alpha}) + p_z(q_{z0} - \frac{p_z}{\alpha})\right) + \right.$$

$$\left. + \frac{1}{2\alpha}\left((q_{x0} - \frac{p_x}{\alpha})^2 + (q_{y0} - \frac{p_y}{\alpha})^2 + (q_{z0} - \frac{p_z}{\alpha})^2\right)\right].$$

Solve (15) for $N_0$

$$N = N_0\left(p_x, p_y, p_z, \frac{p_x}{\alpha} + e^{-\alpha t}(q_x - \frac{p_x}{\alpha}), \frac{p_y}{\alpha} + e^{-\alpha t}(q_y - \frac{p_y}{\alpha}), \frac{p_z}{\alpha} + e^{-\alpha t}(q_z - \frac{p_z}{\alpha})\right).$$

$$\cdot \exp\left\{-k\left[\frac{t}{\alpha^2}(p_x^2 + p_y^2 + p_z^2) + \frac{2}{\alpha^2}(1 - e^{-\alpha t})\left(p_x(q_x - \frac{p_x}{\alpha}) + p_y(q_y - \frac{p_y}{\alpha}) + p_z(q_z - \frac{p_z}{\alpha})\right) + \right.\right. \quad (16)$$

$$\left.\left. + \frac{1}{2\alpha}(1 - e^{-2\alpha t})\left((q_x - \frac{p_x}{\alpha})^2 + (q_x - \frac{p_x}{\alpha})^2 + (q_x - \frac{p_x}{\alpha})^2\right)\right]\right\}.$$

The $N_0(\dots)$ in the (16) means, that one have to calculate $N_0$ from initial density according to (5) and then replace values of its arguments by expressions (11).

Let us specify initial density value as product of delta functions

$$n_0 = \delta(x - x_0)\delta(y - y_0)\delta(z - z_0)\delta(v_x - x_0)\delta(v_y - x_0)\delta(v_z - x_0). \quad (17)$$

The Fourier transform of initial density (17) is

$$N_0 = \frac{1}{(2\pi)^6} \exp(-i(x_0 p_x + y_0 p_y + z_0 p_z + v_{x0} q_x + v_{y0} q_y + v_{z0} q_z)). \quad (18)$$

Substituting (11) for $N_0$ arguments in (18) gives

$$\hat{N}_0 = \frac{1}{(2\pi)^6} \exp\left\{-i\left[x_0 p_x + y_0 p_y + z_0 p_z + \right.\right. \quad (19)$$

$$\left.\left. + v_{x0}\left(\frac{p_x}{\alpha} + e^{-\alpha t}(q_x - \frac{p_x}{\alpha})\right) + v_{y0}\left(\frac{p_y}{\alpha} + e^{-\alpha t}(q_y - \frac{p_y}{\alpha})\right) + v_{z0}\left(\frac{p_z}{\alpha} + e^{-\alpha t}(q_z - \frac{p_z}{\alpha})\right)\right]\right\}$$

or

$$\hat{N}_0 = \frac{1}{(2\pi)^6} \exp\left\{-i\left[p_x\left(x_0 + \frac{v_{x0}}{\alpha}(1 - e^{-\alpha t})\right) + p_y\left(y_0 + \frac{v_{y0}}{\alpha}(1 - e^{-\alpha t})\right) + p_z\left(z_0 + \frac{v_{z0}}{\alpha}(1 - e^{-\alpha t})\right) \right.\right. \quad (20)$$

$$\left.\left. + e^{-\alpha t}(q_x v_{x0} + q_y v_{y0} + q_z v_{z0})\right]\right\}.$$

It is clear that (20) is the Fourier transform of (21)

$$\hat{n}_0 = \delta\left(x - (x_0 + \frac{v_{x0}}{\alpha}(1 - e^{-\alpha t}))\right)\delta\left(y - (y_0 + \frac{v_{y0}}{\alpha}(1 - e^{-\alpha t}))\right)\delta\left(z - (z_0 + \frac{v_{z0}}{\alpha}(1 - e^{-\alpha t}))\right) \quad (21)$$

$$\delta\left(v_x - v_{x0} e^{-\alpha t}\right)\delta\left(v_y - v_{y0} e^{-\alpha t}\right)\delta\left(v_z - v_{z0} e^{-\alpha t}\right).$$

To get inverse Fourier transform of $n$ from its Fourier transform (16) we use two known results (ref. [6]):

1. A product of 2 functions $A(\omega)B(\omega)$ tranforms to convolution $\frac{1}{(2\pi)^n} a(x) * b(x)$ (where $n$ is a number of independent variables).

2. The Gaussian exponent of quadratic form $e^{-\omega^t A \omega}$ with matrix $A$ transforms to exponent of quadratic form with inverse matrix $\sqrt{\frac{\pi}{det(A)}} e^{-\frac{1}{4} x^t A^{-1} x}$.

In our case the matrix is

$$A = k \begin{bmatrix} \frac{t}{\alpha^2} - \frac{2}{\alpha^3}(1 - e^{-\alpha t}) + \frac{1}{2\alpha^3}(1 - e^{-2\alpha t}) & \frac{1}{\alpha^2}(1 - e^{-\alpha t}) - \frac{1}{2\alpha^2}(1 - e^{-2\alpha t}) \\ \frac{1}{\alpha^2}(1 - e^{-\alpha t}) - \frac{1}{2\alpha^2}(1 - e^{-2\alpha t}) & \frac{1}{2\alpha}(1 - e^{-2\alpha t}) \end{bmatrix}. \quad (22)$$

The determinant is equal to

$$det(A) = k^2 \frac{\alpha t(1 - e^{-2\alpha t}) - 2(1 - e^{-\alpha t})^2}{2\alpha^4}. \quad (23)$$

Let us denote by $D$ expression

$$D = \frac{det(A)}{k^2} = \frac{\alpha t(1 - e^{-2\alpha t}) - 2(1 - e^{-\alpha t})^2}{2\alpha^4}. \quad (24)$$

The inverse matrix is

$$A^{-1} = \frac{1}{kD} \begin{bmatrix} \frac{1}{2\alpha}(1 - e^{-2\alpha t}) & -\frac{1}{\alpha^2}(1 - e^{-\alpha t}) + \frac{1}{2\alpha^2}(1 - e^{-2\alpha t}) \\ -\frac{1}{\alpha^2}(1 - e^{-\alpha t}) + \frac{1}{2\alpha^2}(1 - e^{-2\alpha t}) & \frac{t}{\alpha^2} - \frac{2}{\alpha^3}(1 - e^{-\alpha t}) + \frac{1}{2\alpha^3}(1 - e^{-2\alpha t}) \end{bmatrix}. \quad (25)$$

Combining (21) with (25) we obtain expression for the fundamental solution:

$$G = \frac{1}{(2\pi)^6} \left(\frac{\pi}{k\sqrt{D}}\right)^3 \hat{n}_0 * \exp\left\{\frac{-1}{4kD}\left[\frac{1}{2\alpha}(1 - e^{-2\alpha t})(x^2 + y^2 + z^2) - \right.\right. \quad (26)$$

$$-\left(\frac{2}{\alpha^2}(1 - e^{-\alpha t}) - \frac{1}{\alpha^2}(1 - e^{-2\alpha t})\right)(xv_x + yv_y + zv_z) +$$

$$\left.\left.+ \left(\frac{t}{\alpha^2} - \frac{2}{\alpha^3}(1 - e^{-\alpha t}) + \frac{1}{2\alpha^3}(1 - e^{-2\alpha t})\right)(v_x^2 + v_y^2 + v_z^2)\right]\right\};$$

where * means convolution of two functions.

The convolution of arbitrary function with product of delta functions simplifies to substitution delta function arguments for this function arguments. Finally, we obtain

$$G = \frac{1}{(2\pi)^6}\left(\frac{\pi}{k\sqrt{D}}\right)^3 \exp\left\{\frac{-1}{4kD}\left[\frac{1}{2\alpha}(1-e^{-2\alpha t})\left(\hat{x}^2+\hat{y}^2+\hat{z}^2\right)- \right.\right. \qquad (27)$$

$$\left. -\left(\frac{2}{\alpha^2}(1-e^{-\alpha t}) - \frac{1}{\alpha^2}(1-e^{-2\alpha t})\right)\left(\hat{x}\hat{v}_x+\hat{y}\hat{v}_y+\hat{z}\hat{v}_z\right) + \right.$$

$$\left.\left. +\left(\frac{t}{\alpha^2}-\frac{2}{\alpha^3}(1-e^{-\alpha t})+\frac{1}{2\alpha^3}(1-e^{-2\alpha t})\right)\left(\hat{v}_x^2+\hat{v}_y^2+\hat{v}_z^2\right)\right]\right\};$$

where

$$\hat{x} = x - (x_0 + \frac{v_{x0}}{\alpha}(1-e^{-\alpha t})); \hat{y} = y - (y_0 + \frac{v_{y0}}{\alpha}(1-e^{-\alpha t})); \hat{z} = z - (z_0 + \frac{v_{z0}}{\alpha}(1-e^{-\alpha t})); \qquad (28)$$

$$\hat{v}_x = v_x - v_{x0}e^{-\alpha t}; \hat{v}_x = v_y - v_{y0}e^{-\alpha t}; \hat{v}_x = v_z - v_{z0}e^{-\alpha t}.$$

This is the fundamental solution of Fokker - Planck equation.

Let us check validity of solution (27-28). Direct differentiation of (27-28) and substitution to (1) leads to cumbersome calculations. Therefore we use "semi-reverse" method. (27) has Gaussian form, so we search Gaussian solutions of (1):

$$n = \exp\left\{A(t)\left(x^2+y^2+z^2\right) + 2B(t)\left(xv_x+yv_y+zv_z\right) + C(t)\left(v_x^2+v_y^2+v_z^2\right) + 3Q(t)\right\}. \qquad (29)$$

To get rid of exponents we write

$$l = ln(n) = A(t)\left(x^2+y^2+z^2\right) + 2B(t)\left(xv_x+yv_y+zv_z\right) + C(t)\left(v_x^2+v_y^2+v_z^2\right) + 3Q(t). \qquad (30)$$

$l$ must satisfy equation

$$\frac{\partial l}{\partial t} + v_x\frac{\partial l}{\partial x} + v_y\frac{\partial l}{\partial y} + v_z\frac{\partial l}{\partial z} - \alpha\left(v_x\frac{\partial l}{\partial v_x}+v_y\frac{\partial l}{\partial v_y}+v_z\frac{\partial l}{\partial v_z}\right) - 3\alpha = \qquad (31)$$

$$= k\left(\frac{\partial^2 l}{\partial v_x^2}+\frac{\partial^2 l}{\partial v_y^2}+\frac{\partial^2 l}{\partial v_z^2}\right) + k\left(\left(\frac{\partial l}{\partial v_x}\right)^2+\left(\frac{\partial l}{\partial v_x}\right)^2+\left(\frac{\partial l}{\partial v_x}\right)^2\right).$$

instead of equation (1) for $n$.

Substituting (29) for $A, B, C, Q$ in (30) and collecting of similar terms leads to following equtions

$$\frac{dA}{dt} = 4k\ B^2; \qquad (32)$$

$$\frac{dB}{dt} = \alpha B - A + 4k\ B\ C; \qquad (33)$$

$$\frac{dC}{dt} = 2\alpha\ C - 2\ B + 4k\ C^2; \tag{34}$$

$$\frac{dQ}{dt} = \alpha + 2k\ C. \tag{35}$$

It is not easy to solve nonlinear system (32), but our task is simpler. We have only to check, that

$$D(t) = \frac{\alpha t\ (1 - e^{-2\alpha t}) - 2\ (1 - e^{-\alpha t})^2}{2\alpha^4}\ ; \tag{36}$$

$$A(t) = \left(\frac{-1}{4kD}\right)\left(\frac{1}{2\alpha}\ (1 - e^{-2\alpha t})\right); \tag{37}$$

$$B(t) = \left(\frac{-1}{4kD}\right)\left(-\frac{1}{\alpha^2}\ (1 - e^{-\alpha t}) + \frac{1}{2\alpha^2}\ (1 - e^{-2\alpha t})\right); \tag{38}$$

$$C(t) = \left(\frac{-1}{4kD}\right)\left(\frac{t}{\alpha^2} - \frac{2}{\alpha^3}\ (1 - e^{-\alpha t}) + \frac{1}{2\alpha^3}\ (1 - e^{-2\alpha t})\right); \tag{39}$$

$$Q(t) = -\frac{1}{2}\ ln(D(t)). \tag{40}$$

satisfies equations (32-35). Direct substitution proves this.

We proved validity of (27) for the special case

$$x_0 = 0;\ y_0 = 0;\ z_0 = 0;\ v_{x0} = 0;\ v_{y0} = 0;\ v_{z0} = 0. \tag{41}$$

For the common case we prove that differential operators

$$\frac{\partial}{\partial x}\ ;\ \frac{\partial}{\partial y}\ ;\ \frac{\partial}{\partial z}\ ; \tag{42}$$

and

$$\frac{1}{\alpha}\ (1 - e^{-\alpha t})\frac{\partial}{\partial x} + e^{-\alpha t}\frac{\partial}{\partial v_x}\ ;\ \frac{1}{\alpha}\ (1 - e^{-\alpha t})\frac{\partial}{\partial y} + e^{-\alpha t}\frac{\partial}{\partial v_y}\ ;\ \frac{1}{\alpha}\ (1 - e^{-\alpha t})\frac{\partial}{\partial z} + e^{-\alpha t}\frac{\partial}{\partial v_z}\ ; \tag{43}$$

are symmetries of PDE (1). This statement is trivial for (42) because (1) does not contain $x$ explicitly. To prove the statement for (43) we build prolongation of differential operator according to Lie prolongation formula (ref. [7])

$$\delta\left(\frac{\partial u^\alpha}{\partial x^i}\right) = D_i(\delta u^\alpha) - \frac{\partial u^\alpha}{\partial x^k}\ D_i(\delta x^k); \tag{44}$$

where

$u^\alpha$ - a set of dependent variables;

$x^i$ - a set of independent variables;

$D_i$ - full derivation on $x^i$ operator;

$\delta u^\alpha$, $\delta x^k$ - actions of infinitesimal symmetry operator on variables. We use this non-standard notation instead of usual $\zeta^\alpha$, $\xi^i$ to empasize their nature as small variations of variables.

$\delta\left(\frac{\partial u^\alpha}{\partial x^i}\right)$ - induced action of infinitesimal symmetry operator on derivatives.

For operators (43) Lie formula gives

$$\frac{1}{\alpha}(1-e^{-\alpha t})\frac{\partial}{\partial x} - e^{-\alpha t}\frac{\partial}{\partial v_x} + e^{-\alpha t}\left(-\frac{\partial n}{\partial x} + \alpha \frac{\partial n}{\partial v_x}\right)\frac{\partial}{\partial n_t}; \qquad (45)$$

$$\frac{1}{\alpha}(1-e^{-\alpha t})\frac{\partial}{\partial y} - e^{-\alpha t}\frac{\partial}{\partial v_y} + e^{-\alpha t}\left(-\frac{\partial n}{\partial y} + \alpha \frac{\partial n}{\partial v_y}\right)\frac{\partial}{\partial n_t}; \qquad (46)$$

$$\frac{1}{\alpha}(1-e^{-\alpha t})\frac{\partial}{\partial z} - e^{-\alpha t}\frac{\partial}{\partial v_z} + e^{-\alpha t}\left(-\frac{\partial n}{\partial z} + \alpha \frac{\partial n}{\partial v_z}\right)\frac{\partial}{\partial n_t}; \qquad (47)$$

where $n_t = \frac{\partial n}{\partial t}$. For second order derivatives we have $\delta(n_{uu}) = 0$, $\delta(n_{vv}) = 0$, $\delta(n_{ww}) = 0$.

It is easy to calculate, that the action of (45-47) on (1) is identical zero. We proved, that differential operators (42-43) are symmetries of PDE (1).

The action of operators (42-43) on solutions (41) increases variables $x_0, y_0 \cdots$ from zero to arbitrary values. In this way we get from (41) solutions (27-28).

We checked the solution (27-28).

**DISCUSSION**

The main result of these considerations is the closed form expression for fundamental solution of Fokker - Planck equation without forces. This result can be helpful for modelling of more sofisticated problems. It demonstartes in simple and clear way the result of combined diffusion in physical space and velocities space and inertial motion.

---

**REFERENCES**

[1] Stephen Jewson. Weather forecasts, Weather derivatives, Black-Scholes, Feynmann-Kac and Fokker-Planck. arXiv:physics/0312125 v1 20 Dec 2003

[2] George Krylov. On Solvable Potentials for One Dimensional Schrödinger and Fokker-Planck Equations. arXiv:quant-ph/0212046 v1 8 Dec 2002

[3] Alexei Lozinski, Cédric Chauvière. A fast solver for Fokker-Planck equation applied to viscoelastic flows calculations: 2D FENE model, J. Comput. Phys., 189(2) (2003) 607-625.

[4] F. Benamira and L. Guechi. Similarity transformations approach for a generalized Fokker-Planck equation. arXiv:physics/0112002 v1 30 Nov 2001

[5] Luı̌s M. A. Bettencourt. Properties of the Langevin and Fokker-Planck equations for scalar fields and their application to the dynamics of second order phase transitions. arXiv:hep-ph/0005264 v1 25 May 2000

[6] W. Feller, An Introduction to Probability Theory and Its Applications. Volume II, John Wiley and Sons, 1971.

[7] Peter J. Olver, Applications of Lie groups to differential equations. Springer-Verlag, New York, 1986.